\documentclass[]{article}
\usepackage[utf8]{inputenc}
\usepackage{amsmath}
\usepackage[pdftex]{graphicx}
\usepackage{amscd}
\usepackage{calrsfs}
\usepackage{slashed}
\usepackage{amsfonts}
\usepackage{authblk}
\usepackage[letterpaper, portrait, margin=0.7in]{geometry}
\usepackage{float}
\usepackage{caption}
\usepackage{subcaption}
\title{Quantum propagators for geodesic congruences}
\author{Miguel Socolovsky}
\affil{Instituto de Ciencias Nucleares, Universidad Nacional Aut\'onoma de M\'exico, Cd. Universitaria, 04510, Ciudad de M\'exico, M\'exico\\socolovs@nucleares.unam.mx}

\providecommand{\keywords}[1]
{
	\small	
	\textbf{Keywords:} #1
}
\begin{document}
\date{}
\maketitle

\begin{abstract}
Using the Raychaudhuri equation, we show that a quantum probability amplitude (Feynman propagator) can be univocally associated to any timelike or null affinely parametrized geodesic congruence.	
\end{abstract}

\keywords{Raychaudhuri equation; geodesic congruence; Feynman propagator; Schwarzschild metric}

\section{Introduction}

The Raychaudhuri equation [1] is an entirely geometrical equation, in the sense that it provides a description of the evolution or flow of congruences of timelike or lightlike (null) curves, these been geodesics or with an acceleration term, in a given spacetime, that is, a differentiable manifold equipped with a pseudo-Riemannian metric $g_{\mu\nu}$ inducing the Levi-Civita connection $\Gamma^\mu_{\nu\rho}$. The characteristic elements of the congruences are the expansion scalar $\Theta$ and the shear and rotation tensors, respectively $\sigma_{\mu\nu}$ and $\omega_{\mu\nu}$, for geodesic curves, plus an acceleration term $a^\mu$ in the non-geodesic case. The information about the geometry of the spacetime lies in the Ricci tensor $R_{\mu\nu}$ determined by the metric. The connection with physics appears due to the relation of these two tensors with the energy-momentum tensor $T_{\mu\nu}$ through the Einstein equation $R_{\mu\nu}-{{1}\over{2}}g_{\mu\nu}R-\Lambda g_{\mu\nu}=8\pi T_{\mu\nu}$, where $R$ is the curvature scalar $R^\mu_\mu$ and $\Lambda$ is the cosmological constant.

For geodesics, the simplest form of the Raychaudhuri equation is when they are affinely parametrized, that is with a parameter $\lambda$ uniquely determined up to an affine transformation $\lambda\to\lambda^\prime=a\lambda+d$, $a,d\in\mathbb{R}$, $a\neq 0$. 

What constitutes the crucial point of the present analysis, is that with the transformation indicated in equation (7) below (Section 3) [2,3], the Raychaudhuri equation reduces to a 1-dimensional {\it harmonic oscillator equation with a ``time" ($\lambda$)-dependent frequency} $\Omega$, which, being a second order homogeneous ordinary differential equation for the function $F(\lambda)$, can be integrated with a suitable domain of definition of the parameter and adequate boundary conditions. Since in general $\Omega$ is not a periodic function of its argument, (8) is not a Hill equation [4], but is known as a ``Hill-type" equation. 

Once a $\lambda$-dependent Lagrangian leading to the oscillator equation is defined, a Feynman path integral [5,6] leads to a propagator $K(F^{\prime\prime},\lambda^{\prime\prime};F^\prime,\lambda^\prime)$ from given initial $(F^\prime,\lambda^\prime)$ to final $(F^{\prime\prime},\lambda^{\prime\prime})$ values of the affine parameter and of the function $F$ representing the expansion (Section 4). The propagator is essentially a quantum object since it is obtained by a functional integration over {\it all} fluctuations of the expansion -represented by $F$- along its classical evolution; so $K$ is nothing but the quantum description of the congruence flow. In contrast with the classical case where the expansions diverge at a caustic or at a singularity of the metric (e.g. at the future and past singularities $r_\pm\equiv 0\vert_\pm$ in the Schwarzschild-Kruskal-Szekeres black hole [7,8]), the associated propagators remain finite. The motivation of this ``quantization" of the Raychaudhuri equation is to explore the idea that the non spacelike geodesics are the {\it fundamental quantities of gravity theory}, perhaps even more fundamental than the metric itself [9]. This conception is related to the loop formulation of quantum gravity based on the ideas that forces are described by lines (e.g. Wilson loops) and the notion of background independence (spacetime/gravitational field identification) [10]. For a recent introductory review of other approaches to quantum gravity see, e.g., Ref. [11].

In Section 5 we apply the above general construction and obtain the propagators of simple but non-trivial ingoing and outgoing radial timelike geodesic congruences outside the horizon of the Schwarzschild black hole (Subsection 5.1) and of radial null ingoing and outgoing geodesics respectively in the black hole and white hole regions of this metric (Subsection 5.2). Section 6 is devoted to final comments.
\section{Raychaudhuri equation}
Let $v=(v^\mu)$, $v^\mu={{dx^\mu}\over{d\lambda}}$, be the vector field tangent to an affinely parametrized timelike (T.L.) or null (N) geodesic congruence ($\lambda$ is the affine parameter) in a 4-dimensional spacetime with local coordinates $x^\mu$, $\mu=0,1,2,3$, metric $g_{\mu\nu}$, Levi-Civita connection 
$\Gamma^\mu_{\nu\rho}={{1}\over{2}}g^{\mu\sigma}(\partial_\nu g_{\rho\sigma}+\partial_\rho g_{\nu\sigma}-\partial_\sigma g_{\nu\rho})$, curvature tensor $R^\rho_{\mu\sigma\nu}=\Gamma^\rho_{\mu\nu,\sigma}-\Gamma^\rho_{\mu\sigma,\nu}+\Gamma^\lambda_{\mu\nu}\Gamma^\rho_{\lambda\sigma}-\Gamma^\lambda_{\mu\sigma}\Gamma^\rho_{\lambda\nu}$, Ricci tensor $R_{\mu\nu}=R^\rho_{\mu\rho\nu}$, and covariant derivative $D=(D_\mu)$. $v$ obeys the equation 
\begin{equation}
v\cdot D(v^\mu)=v^\nu D_\nu(v^\mu)=v^\nu(\partial_\nu v^\mu+\Gamma^\mu_{\nu\rho}v^\rho)=0
\end{equation}
with normalization $v^2=g_{\mu\nu}v^\mu v^\nu=+1$ (0) in the T.L.  (N) case (we use the signature (+,-,-,-)). The {\it expansion} of the congruence, that is, the fractional rate of change of the cross-sectional volume (area) to the congruence in the T.L. (N) case, is the scalar 
\begin{equation}
\Theta=D\cdot v={{1}\over{\sqrt{-g}}}\partial_\mu(\sqrt{-g}v^\mu),
\end{equation} 
where $g=det(g_{\mu\nu})$. Through pure geometrical identities, $\Theta$ can be shown to obey the Raychaudhuri equation [1,2] (a Riccati equation)
\begin{equation}
{{d\Theta}\over{d\lambda}}=-{{1}\over{n}}\Theta^2-\sigma_{\mu\nu}\sigma^{\mu\nu}+\omega_{\mu\nu}\omega^{\mu\nu}-R_{\mu\nu}v^\mu v^\nu
\end{equation}
i.e. $\dot{\Theta}=-{{1}\over{n}}\Theta^2-\sigma^2+\omega^2-Ric(v,v)$, where $n=3$ (2) in the T.L. (N) case; $\sigma_{\mu\nu}$ (shear, which measures the change in shape of the congruence without modification of its volume in the T.L. case, or of its area in the N case) and $\omega_{\mu\nu}$ (rotation) are, respectively, the traceless symmetric and antisymmetric parts of the tensor 
\begin{equation}
B_{\mu\nu}=D_\nu v_\mu,
\end{equation}
so that $\Theta=B^\mu_{;\mu}$. One has the decomposition
\begin{equation}
B_{\mu\nu}=\sigma_{\mu\nu}+\omega_{\mu\nu}+{{1}\over{n}}\Theta h_{\mu\nu},
\end{equation}
where $h_{\mu\nu}$ is the transverse metric (part of $g_{\mu\nu}$ orthogonal to $v$) given by $g_{\mu\nu}-v_\mu v_\nu$ in the T.L. case and $g_{\mu\nu}-(v_\mu n_\nu+v_\nu n_\mu)$ in the N case ($n^\mu$ is a null vector satisfying $v\cdot n=+1$). 

\

(3) is a purely geometrical equation; its physical meaning [12]  only comes after relating the Ricci tensor to the energy-momentum tensor $T_{\mu\nu}$ through the Einstein equation 
\begin{equation}
R_{\mu\nu}-{{1}\over{2}}g_{\mu\nu}R-\Lambda g_{\mu\nu}=8\pi T_{\mu\nu},
\end{equation}
where $R=R^\mu_\mu$ and $\Lambda$ is the cosmological constant; finally, all the terms in (3) depend on $\lambda$ through the $x^\mu$'s. In the vacuum, $T_{\mu\nu}=\Lambda=0$, implying $R_{\mu\nu}=0$.

\

(Units: In the geometrical system, $G=c=1$, so if $[\lambda]=[L]$, then $[\Theta]=[L]^{-1}$, $[\sigma]^2=[\omega]^2=[R_{\mu\nu}]=[L]^{-2}$, and $[v]=[L]^0$.)

\

\section{Frequency dependent harmonic oscillator}

In terms of the function $F(\lambda)$ defined by [2,3] 
\begin{equation}
\Theta(\lambda)=n{{\dot{F}(\lambda)}\over{F(\lambda)}},
\end{equation}
the Raychoudhuri equation (3) becomes 
\begin{equation}
\ddot{F}(\lambda)+(\Omega(\lambda))^2F(\lambda)=0
\end{equation}
with 
\begin{equation}
\Omega^2={{1}\over{n}}(\sigma^2-\omega^2+R_{\mu\nu}v^\mu v^\nu),
\end{equation}
which is nothing but the equation of a classical 1-dimensional harmonic oscillator with $\lambda$ (``time")-dependent frequency $\Omega$. After Hill [4], (8) is known as a ``Hill-type" equation. If at $\lambda=\lambda_0$ the congruence converges to a point i.e. $\Theta(\lambda)$ has a caustic: $\Theta(\lambda)\to -\infty$ as $\lambda \to \lambda_0$, then $\lambda_0$ must be a zero of $F(\lambda)$ if $\dot{F}(\lambda_0)$ is finite.

\

(8) is the Euler-Lagrange equation of the ``time"-dependent Lagrangian 
\begin{equation}
{\cal L}(F,\dot{F},\lambda)={{1}\over{2}}(\dot{F}^2-\Omega^2F^2).
\end{equation}
For a suitable domain of definition of $\lambda$, (8) admits a solution $\bar{F}(\lambda)$ subject to the boundary conditions $F^\prime=\bar{F}(\lambda^\prime)$ and $F^{\prime\prime}=\bar{F}(\lambda^{\prime\prime})$ with, e.g., $\lambda^\prime<\lambda^{\prime\prime}$.

\

(Units: [$F$]=$[L]^{1/2}$ since [action]=$[\int d\lambda{\cal L}]=[L][{\cal L}]=[L]^0.$) 
\section{Path integrals and time-dependent quadratic Lagrangians}

It is well known [5,6] that a Lagrangian of the form 
\begin{equation}
{\cal L}(x,\dot{x},t)={{1}\over{2}}((\dot{x}(t))^2-b(t)(x(t))^2)
\end{equation}
has associated with it a perfectly defined propagator $K(x^{\prime\prime},t^{\prime\prime};x^\prime,t^\prime)$ from the quantum state $\vert x^\prime,t^\prime>$ to the quantum state $\vert x^{\prime\prime},t^{\prime\prime}>$ given by the path integral 
\begin{equation}
\int_{x(t^\prime)=x^\prime}^{x(t^{\prime\prime})=x^{\prime\prime}}{\cal D}x(t)e^{i\int_{t^\prime}^{t^{\prime\prime}}dt{\cal L}(x,\dot{x},t)},
\end{equation}
($\hbar=1$) where, formally,
\begin{equation}
\int_{x(t^\prime)=x^\prime}^{x(t^{\prime\prime})=x^{\prime\prime}}{\cal D}x(t)...=\prod_{t\in(t^\prime,t^{\prime\prime})}\int_{-\infty}^{+\infty}dx(t)... \ . 
\end{equation}
The result is
\begin{equation}
K(x^{\prime\prime},t^{\prime\prime};x^\prime,t^\prime)=(2\pi if(t^{\prime\prime},t^\prime))^{-1/2}e^{iS[\bar{x}]},
\end{equation}
where $\bar{x}(t)$ is the solution of 
\begin{equation}
\ddot{x}(t)+b(t)x(t)=0
\end{equation}
with $x(t^{\prime\prime})=x^{\prime\prime}$ and $x(t^\prime)=x^\prime$,
\begin{equation}
S[\bar{x}]=\int_{t^\prime}^{t^{\prime\prime}}dt{\cal L}(\bar{x}(t),\dot{\bar{x}}(t),t),
\end{equation}
and $f(t,t^\prime)$ is the solution of 
\begin{equation}
{{\partial^2f(t,t^\prime)}\over{\partial t^2}}+b(t)f(t,t^\prime)=0
\end{equation}
with $f(t^\prime,t^\prime)=0$ and ${{\partial f(t,t^\prime)}\over{\partial t}}\vert_{t=t^\prime}=0$.

\

Since (10) and (11) (and therefore (8) and (15)) have the same form, then 
\begin{equation}
K(F^{\prime\prime},\lambda^{\prime\prime};F^\prime,\lambda^\prime)=
\int_{F(\lambda^\prime)=F^\prime}^{F(\lambda^{\prime\prime})=F^{\prime\prime}}{\cal D}F(\lambda)e^{i\int_{\lambda^\prime}^{\lambda^{\prime\prime}}d\lambda{\cal L}(F,\dot{F},\lambda)}=(\prod_{\lambda\in(\lambda^\prime,\lambda^{\prime\prime})}\int_{-\infty}^{+\infty}dF(\lambda))e^{i\int_{\lambda^\prime}^{\lambda^{\prime\prime}}d\lambda{\cal L}(F,\dot{F},\lambda)}=(2\pi if(\lambda^{\prime\prime},\lambda^\prime))^{-1/2}e^{iS[\bar{F}]}
\end{equation} 
with
\begin{equation}
S[\bar{F}]=\int_{\lambda^\prime}^{\lambda^{\prime\prime}}d\lambda{\cal L}(\bar{F}(\lambda),\dot{\bar{F}}(\lambda),\lambda) 
\end{equation}
and $f(\lambda,\lambda^\prime)$ solution of (17) with $t$'s replaced by $\lambda$'s, is a Feynman propagator and therefore a quantum object describing the flow of the geodesic congruence from $\lambda=\lambda^\prime$ to $\lambda=\lambda^{\prime\prime}$. To the pairs $(F^{\prime\prime},\lambda^{\prime\prime})$ and $(F^\prime,\lambda^\prime)$ might correspond ``quantum states" $\vert F^{\prime\prime},\lambda^{\prime\prime}>$ and $\vert F^\prime,\lambda^\prime>$, belonging to the Hilbert space of a loop quantum gravity theory, as superpositions of an orthonormal basis of spin network states, which in turn are linear combinations of loop states [10].
\section{Examples}
\subsection*{5.1}
As a first example of a quantum propagator associated to a geodesic flow, we consider the outgoing (+ sign) and ingoing (- sign) timelike radial geodesic congruences given by equation (2) in Ref. [2] in the unique spherically symmetric asymptotically flat vacuum ($T_{\mu\nu}=\Lambda=0$) solution of (6), namely the Schwarzschild spacetime, with metric 
\begin{equation}
ds^2=fdt^2-f^{-1}dr^2-r^2d\Omega_2, \ d\Omega_2=d\theta^2+sin^2\theta d\varphi^2,
\end{equation}
\begin{equation}
g_{tt}=f=1-{{2M}\over{r}}, \ g_{rr}=-f^{-1}, \ g_{\theta\theta}=-r^2, \ g_{\varphi,\varphi}=-r^2sin^2\theta, \  x^\mu=(t,r,\theta,\varphi), \ t\in(-\infty,+\infty), \ r>2M, \ \theta\in[0,\pi], \ \varphi\in[0,2\pi).
\end{equation}

The tangent vector fields to the geodesics are given by
\begin{equation}
v_\pm=v_\pm^\mu\partial_\mu=v_\pm^t\partial_t+v_\pm^r\partial_r={{1}
\over{f}}\partial_t\pm\sqrt{{{2M}\over{r}}}\partial_r
\end{equation} 
i.e. $v_\pm^t={{1}\over{f}}$, $v_\pm^r=\pm\sqrt{{{2M}\over{r}}}$, $v_\pm^\theta=v_\pm^\varphi=0$. 

\

It is easily verified that the geodesics are {\it affinely parametrized}:
\begin{equation}
v_\pm\cdot D(v_\pm^\mu)=0
\end{equation}
and {\it hypersurface orthogonal}: in fact,
\begin{equation}
v_{\pm t}=g_{tt}v_\pm^t=\partial_t\Phi_\pm(t,r), \ v_{\pm r}=g_{rr}v_\pm^r=\partial_r\Phi_\pm(t,r) 
\end{equation} 
with hypersurfaces defined by
\begin{equation}
\Phi(t,r)_\pm=\mp\int^rdr^\prime{{\sqrt{{{2M}\over{r^\prime}}}\over{1-{{2M}\over{r^\prime}}}}}+t=\mp 4M({{1}\over{z}}-{{1}\over{2}}ln({{1+z}\over{1-z}}))+t=const.,
\end{equation}
where we used 2.149 and 2.172 of Ref. [13], and $z=\sqrt{{{2M}\over{r}}}$. 
As a consequence, the geodesics {\it rotation vanishes} [15]: 
\begin{equation}
\omega_{\pm\mu\nu}=0. 
\end{equation}
From the {\it non-vanishing} components of the {\it shear} tensors, given in equations (4) and (5) in Ref. [2], and using the inverse metric $g^{\mu\nu}=diag(f^{-1},-f,-r^{-2},-r^{-2}sin^{-2}\theta)$, a straightforward calculation leads to 
\begin{equation}
\sigma_\pm^2=\sigma_{\pm\mu\nu}\sigma^{\mu\nu}_\pm={{3M}\over{r^3}}.
\end{equation}
Equations (2) and (22) lead to the expansions
\begin{equation}
\Theta_\pm=\pm{{3}\over{2}}\sqrt{{{2M}\over{r^3}}}
\end{equation} 
and therefore to the fulfillment of the Raychoudhuri equations
\begin{equation}
{{d\Theta_\pm}\over{d\lambda_\pm}}+{{1}\over{3}}\Theta_\pm^2+\sigma_\pm^2=0
\end{equation}
with affine parameters obeying 
\begin{equation}
{{dr}\over{d\lambda_\pm}}=\pm\sqrt{{{2M}\over{r}}}=v_\pm^r.
\end{equation}
Up to an additive constant, one obtains
\begin{equation}
\lambda_\pm=\pm{{1}\over{3}}\sqrt{{{2r^3}\over{M}}},
\end{equation}
with
\begin{equation}
\lambda_+\in({{4M}\over{3}},+\infty)
\end{equation}
and 
\begin{equation}
\lambda_-\in(-\infty,-{{4M}\over{3}}).
\end{equation}
In terms of $\lambda_\pm$,
\begin{equation}
\Theta_\pm={{1}\over{\lambda_\pm}}, \ {{d\Theta_\pm}\over{d\lambda_\pm}}=-{{1}\over{\lambda_\pm^2}}, \ \sigma_\pm^2={{2}\over{3\lambda_\pm^2}},
\end{equation}
which fulfills (29). Also, for the flow of both the future directed outgoing (from $\lambda^\prime_+$ to $\lambda^{\prime\prime}_+$) and ingoing (from $\lambda^\prime_-$ to $\lambda^{\prime\prime}_-$) congruences, $\lambda^\prime_\pm\leq\lambda_\pm\leq\lambda^{\prime\prime}_\pm$. In particular, it can be easily seen that if $\Theta_-$ is negative at $(\lambda_-)_0$, the ingoing congruence converges to a point in a finite interval of the affine parameter ({\it focusing theorem}) i.e. it has a {\it caustic}: from the fact that $\sigma_-^2={{2}\over{3\lambda_-^2}}>0$, ${{d\Theta_-}\over{d\lambda_-}}<-{{1}\over{3}}\Theta_-^2$ which implies 
$\int^{\Theta_-}_{(\Theta_-)_0}{{d\Theta^\prime_-}\over{(\Theta^\prime_-)^2}}=-{{1}\over{\Theta_-(\lambda_-)}}+{{1}\over{(\Theta_-)_0}}<{{1}\over{3}}(\lambda_--(\lambda_-)_0)$ where $(\Theta_-)_0=\Theta_-((\lambda_-)_0)$. Then ${{1}\over{\Theta_-(\lambda_-)}}>-{{1}\over{\vert(\Theta_-)_0\vert}}+{{1}\over{3}}(\lambda_--(\lambda_-)_0)$ since $(\Theta_-)_0<0$. So, if $(\lambda_--(\lambda_-)_0)\to ({{3}\over{\vert(\Theta_-)_0\vert}})_-$, then
\begin{equation}
{{1}\over{\Theta_-(\lambda_-)}}\to 0_- \ \Longleftrightarrow \ \Theta_-(\lambda_-)\to-\infty.
\end{equation}  
From $v_\pm^t={{dt}\over{d\lambda_\pm}}={{1}\over{{1-{{2M}\over{r}}}}}=\pm \sqrt{{{2M}\over{r}}}({{dt}\over{dr}})_\pm$ one obtains
\begin{equation}
({{dt}\over{dr}})_\pm=\pm{{1}\over{\sqrt{{{2M}\over{r}}}(1-{{2M}\over{r}})}}
\end{equation}
from which it follows 
\begin{equation}
(t(r,r_0))_\pm=\pm\int_{r_0}^r{{dr^\prime}\over{(2M/r^\prime)^{1/2}-(2M/r^\prime)^{3/2}}}=\pm 4M({{1}\over{3}}(z^{-3/2}-z_0^{-3/2})+(z^{-1/2}-z_0^{-1/2}))-{{1}\over{2}}ln({{(1+z)(1-z_0)}\over{(1-z)(1+z_0)}})),
\end{equation}
where again we used 2.149 and 2.172 of Ref. [13], $2M<r_0,r$, $r_0<r$ (outgoing case), $r<r_0$ (ingoing case), $z_0=\sqrt{{{2M}\over{r_0}}}$, and $z=\sqrt{{{2M}\over{r}}}$.

\

Defining the functions $F_\pm(\lambda_\pm)$ through 
\begin{equation}
\Theta_\pm(\lambda_\pm)=3{{\dot{F}_\pm(\lambda_\pm)\over{F_\pm(\lambda_\pm)}}}
\end{equation}
the Raychoudhuri equations (29) become
\begin{equation}
\ddot{F}_\pm(\lambda_\pm)+\Omega^2_\pm(\lambda_\pm)F_\pm(\lambda_\pm)=0
\end{equation} 
with affine parameter-dependent square frequency
\begin{equation}
\Omega^2_\pm(\lambda_\pm)={{2}\over{9\lambda^2_\pm}}.
\end{equation}
Since $\lambda_\pm^2\in({{16}\over{9}}M^2,+\infty)$, then $\Omega_\pm\in(0,{{1}\over{2\sqrt{2}M}})$.

\

Using the series (or Frobenius) expansion method for obtaining solutions of linear second order homogeneous ordinary differential equations [14], the solutions $\bar{F}_\pm(\lambda_\pm)$ of (39) with the boundary conditions 
\begin{equation}
F^\prime_\pm=\bar{F}_\pm(\lambda^\prime_\pm), \ F^{\prime\prime}_\pm=\bar{F}_\pm(\lambda^{\prime\prime}_\pm)
\end{equation}
are
\begin{equation}
\bar{F}_\pm(\lambda_\pm)={{(F^\prime_\pm\lambda_\pm^{\prime\prime \ 1/3}-F^{\prime\prime}_\pm\lambda_\pm^{\prime \ 1/3})\lambda_\pm^{2/3}+(F^{\prime\prime}_\pm\lambda_\pm^{\prime \ 2/3}-F^\prime_\pm\lambda_\pm^{\prime\prime \ 2/3})\lambda_\pm^{1/3}}\over{(\lambda_\pm^{\prime\prime}\lambda_\pm^\prime)^{1/3}((\lambda_\pm^\prime)^{1/3}-(\lambda_\pm^{\prime\prime})^{1/3}) }}.
\end{equation}
The associated quantum propagators from the ``quantum states" $\vert F^\prime_\pm,\lambda^\prime_\pm>$ to the ``quantum states"  $\vert F^{\prime\prime}_\pm,\lambda^{\prime\prime}_\pm>$ are obtained from the results in section 4 with the replacements $t\to\lambda_\pm$, $\bar{x}(t)\to\bar{F}_\pm(\lambda_\pm)$, $a(t)\to \Omega^2_\pm(\lambda_\pm)$, and by explicitly solving (17) with the above changes. The result is:

\begin{equation}
K_\pm(F_\pm^{\prime\prime},\lambda_\pm^{\prime\prime};F_\pm^\prime,\lambda_\pm^\prime)=(2\pi i f(\lambda_\pm^{\prime\prime},\lambda_\pm^\prime))^{-1/2}e^{iS[\bar{F}_\pm]},
\end{equation}
with
\begin{equation}
S[\bar{F}_\pm]={{1}\over{2}}\int_{\lambda_\pm^\prime}^{\lambda_\pm^{\prime\prime}}d\lambda_\pm((\dot{\bar{F}}_\pm(\lambda_\pm))^2-{{2}\over{9\lambda_\pm^2}}(\bar{F}_\pm(\lambda_\pm))^2) 
\end{equation}
given by
\begin{equation}
{{(F^\prime_\pm)^2\lambda^{\prime\prime}_\pm(1-{{1}\over{2}}((\lambda^\prime_\pm/\lambda^{\prime\prime}_\pm)^{1/3}+(\lambda^{\prime\prime}_\pm/\lambda^\prime_\pm)^{1/3}))- (F^{\prime\prime}_\pm)^2\lambda^\prime_\pm(1-{{1}\over{2}}((\lambda^{\prime\prime}_\pm/\lambda^\prime_\pm)^{1/3}+(\lambda^\prime_\pm/\lambda^{\prime\prime}_\pm)^{1/3}))}\over{3(\lambda^{\prime\prime}_\pm\lambda^\prime_\pm)^{2/3}((\lambda^\prime_\pm)^{1/3}-(\lambda^{\prime\prime}_\pm)^{1/3})^2}}, 
\end{equation}
and
\begin{equation}
f(\lambda_\pm^{\prime\prime},\lambda_\pm^\prime)=3(({{\lambda_\pm^\prime}\over{\lambda_\pm^{\prime\prime}}})^{1/3}-({{\lambda_\pm^\prime}\over{\lambda_\pm^{\prime\prime}}})^{2/3}).
\end{equation}
In the path integrals defining $K_\pm(F_\pm^{\prime\prime},\lambda_\pm^{\prime\prime};F_\pm^\prime,\lambda_\pm^\prime)$, the ``integration measures" for the (+) and (-) cases are 
\begin{equation}
\prod_{\lambda_\pm\in(\lambda_\pm^\prime,\lambda_\pm^{\prime\prime})}\int_{-\infty}^{+\infty}dF_\pm(\lambda_\pm)... 
\end{equation}
with, respectively, $(\lambda^\prime_+,\lambda^{\prime\prime}_+)\subset ({{4M}\over{3}},+\infty)$ and $(\lambda^\prime_-,\lambda^{\prime\prime}_-)\subset (-\infty,-{{4M}\over{3}})$. 

\

 For the ingoing congruence, we saw in the previous subsection that if at $\lambda^\prime_-=(\lambda^\prime_-)_0$ the expansion $(\Theta_-)_0$ is negative, then it diverges at $\lambda^{\prime\prime}=(\lambda^\prime_-)_0+{{3}\over{\vert (\Theta_-)_0\vert}}$ where, because of (7), $F^{\prime\prime}_-=\bar{F}(\lambda^{\prime\prime}_-)=0$. However, in contradistinction with this divergence, according to (43), (45) and (46), the propagator from $(F^\prime,(\lambda^\prime_-)_0)$ to $(0,\lambda^{\prime\prime}_-)$ remains {\it finite} and is given by 
\begin{equation}
K_-(0,\lambda^{\prime\prime}_-;F^\prime_-,(\lambda^\prime_-)_0)={{e^{i(S[\bar{F}_-])+\pi/4)}}\over{\sqrt{6\pi(((\lambda^\prime_-)_0/\lambda^{\prime\prime}_-)^{2/3}-((\lambda^\prime_-)_0/\lambda^{\prime\prime}_-)^{1/3})}}}    
\end{equation}
with
\begin{equation}
S[\bar{F}_-]={{(F^\prime_-)^2\lambda^{\prime\prime}_-(1-{{1}\over{2}}(((\lambda^\prime_-)_0/\lambda^{\prime\prime}_-)^{1/3}+(\lambda^{\prime\prime}_-/(\lambda^\prime_-)_0)^{1/3}))} \over{3(\lambda^{\prime\prime}_-(\lambda^\prime_-)_0)^{2/3}(((\lambda^\prime_-)_0)^{1/3}-(\lambda^{\prime\prime}_-)^{1/3})^2}}.
\end{equation}
(48) gives the wave function $\psi(0,\lambda^{\prime\prime}_-)$  associated to the ket $\vert 0,\lambda^{\prime\prime}_->$, if the wave function associated with the ket $\vert F^\prime_-,(\lambda^\prime_-)_0>$ is $\delta(\lambda^\prime_--(\lambda^\prime_-)_0)$:
\begin{equation}
\psi(0,\lambda^{\prime\prime}_-)=\int_{-\infty}^{-4M/3}d\lambda^\prime K_-(0,\lambda^{\prime\prime}_-;F^\prime_-,\lambda^\prime_-)\delta(\lambda^\prime_--(\lambda^\prime_-)_0)=K_-(0,\lambda^{\prime\prime}_-;F^\prime,(\lambda^\prime_-)_0). 
\end{equation}
\subsection*{5.2}
As a second example, we consider the propagators associated to radial future directed null geodesics in the interior of the black hole (B.H.) and white hole (W.H.) regions of the Schwarzschild-Kruskal-Szekeres metric respectively starting at the future and past horizons $H^+$ and $H^-$ and ending in the future singularity at $r=0\vert_+$ (ingoing geodesics, cases (a) and (b) in Figure 1), and starting in the past singularity at $r=0\vert_-$ and ending in the past and future horizons $H^-$ and $H^+$ (outgoing geodesics, cases (c) and (d) in Figure 2).

\begin{figure}[H]
	\centering
	\begin{subfigure}{.5\textwidth}
		\centering
		\includegraphics[width=.7\linewidth]{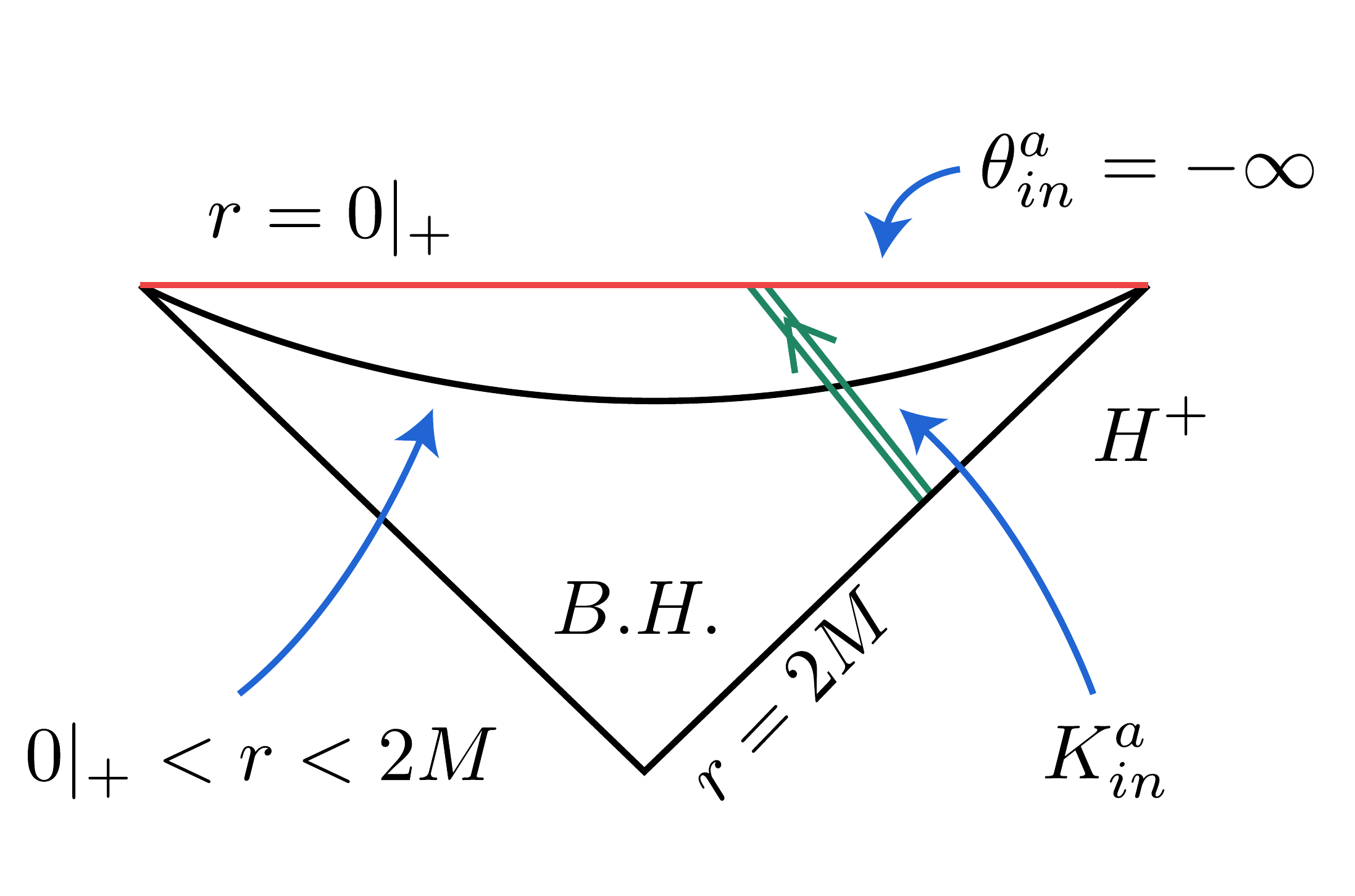}
		\caption{}
		\label{fig:sub1a}
	\end{subfigure}%
	\begin{subfigure}{.5\textwidth}
		\centering
		\includegraphics[width=.7\linewidth]{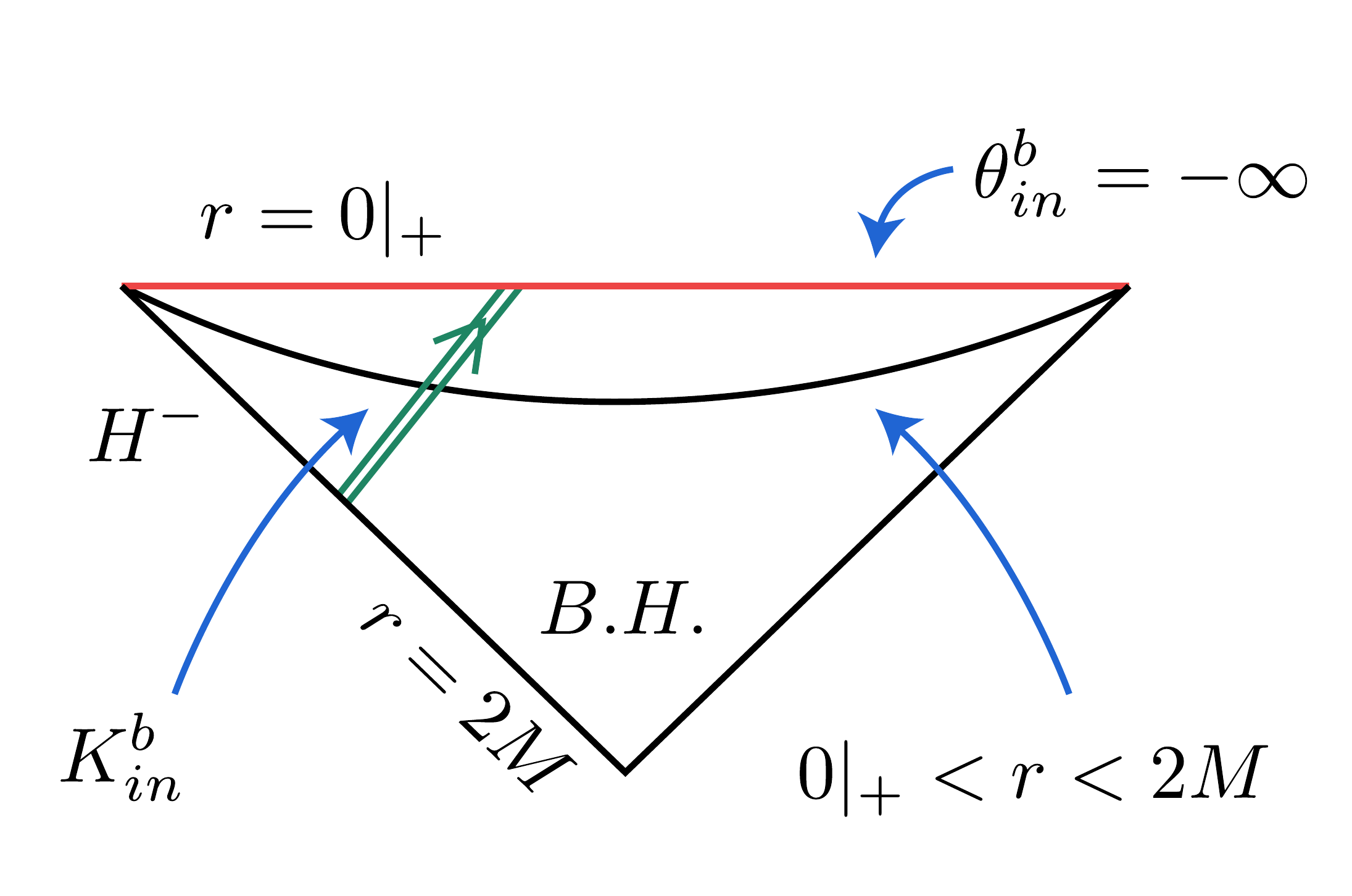}
		\caption{}
		\label{fig:sub1b}
	\end{subfigure}
	\caption{Future directed null ingoing geodesics propagators: (a): $H^+\to 0\vert_+$, (b): $H^-\to =\vert_+$}
	\label{fig1}
\end{figure}

\begin{figure}[H]
	\centering
	\begin{subfigure}{.5\textwidth}
		\setcounter{subfigure}{2}
		\centering
		\includegraphics[width=.7\linewidth]{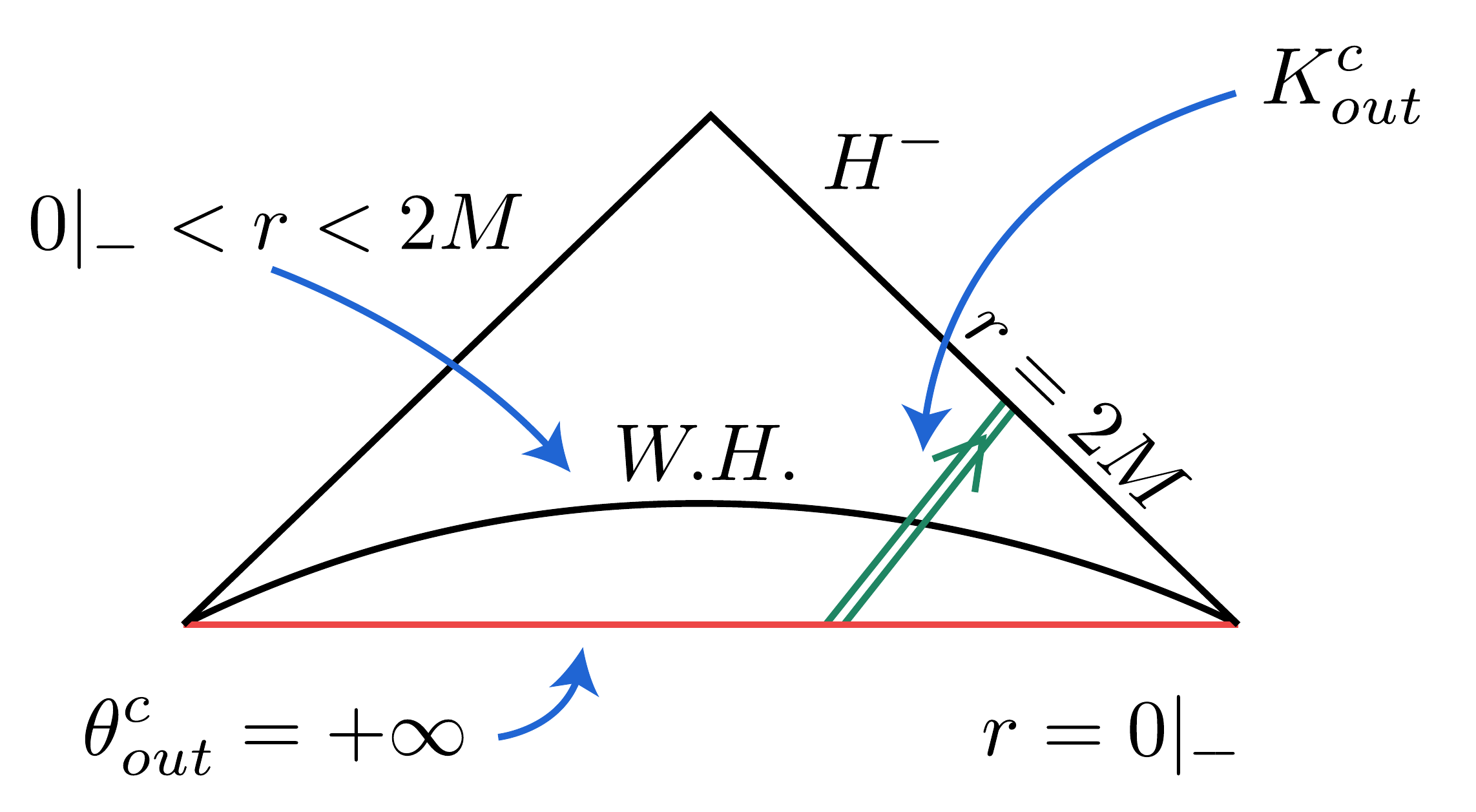}
		\caption{}
		\label{fig:sub2a}
	\end{subfigure}%
	\begin{subfigure}{.5\textwidth}
		\centering
		\includegraphics[width=.7\linewidth]{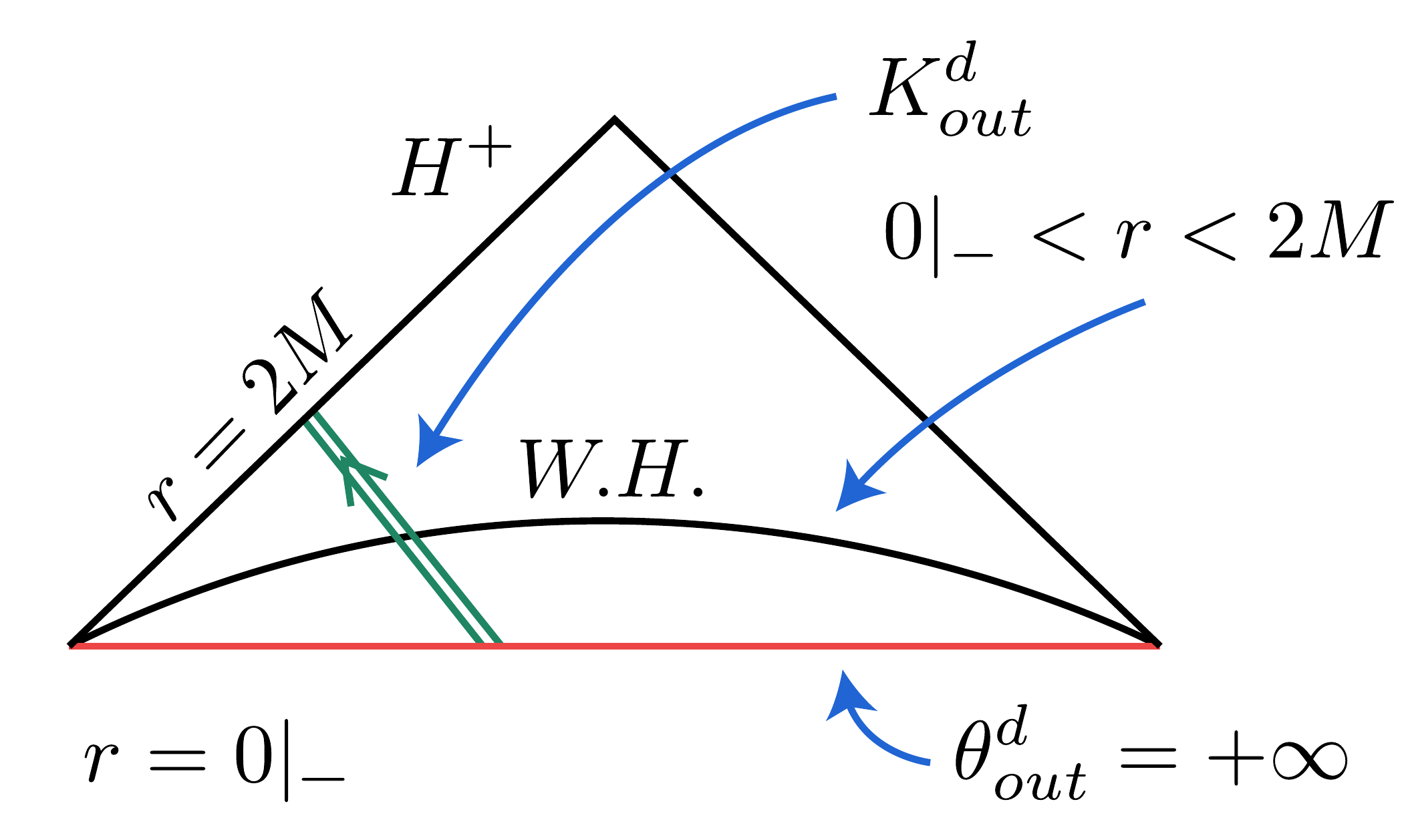}
		\caption{}
		\label{fig:sub2b} 
	\end{subfigure}
	\caption{Future directed null outgoing geodesics propagators: (c): $0\vert-\to H^-$, (d): $0\vert_-\to H^+$}
	\label{fig2}
\end{figure}

As is well known [15,16], the expansions are given by
\begin{equation}
\Theta_{in}^a=\Theta_{in}^b=-{{2}\over{r}}\in(-\infty,-1/M)
\end{equation}  
and
\begin{equation}
\Theta_{out}^c=\Theta_{out}^d=+{{2}\over{r}}\in(1/M,+\infty)
\end{equation}
with vanishing rotation ($\omega_{\mu\nu}=0$), shear ($\sigma_{\mu\nu}=0$), and Ricci tensor ($R_{\mu\nu=0}$). The afine parameters are $\lambda=\mp r$, respectively for the ingoing and the outgoing cases, and the Raychaudhuri equations (3) are
\begin{equation}
{{d\Theta}\over{d\lambda}}=-{{1}\over{2}}\Theta^2
\end{equation} 
for $\Theta^x_{in}$, $x=a,b$; $\lambda=-r$, and $\Theta^y_{out}$, $y=c,d$; $\lambda=+r$. Defining $F$ as in (7) with $n=2$, since $\Omega=0$ one obtains 
\begin{equation}
\ddot{F}(\lambda)=0
\end{equation}
(free non relativistic particle!) with solution 
\begin{equation}
F(\lambda)=A\lambda+B.
\end{equation}
Since as $\lambda=\mp r\to 0\vert_\pm$, $\Theta^x_{in}\to-\infty$ and $\Theta^y_{out}\to+\infty$, the corresponding $F's\to 0$ and so $B=0$ for all cases. Then
\begin{equation}
F^x_{in}(r)={{F^x_{in2}-F^x_{in1}}\over{r_2-r_1}}r, \ F^y_{out}(r)={{F^y_{out2}-F^y_{out1}}\over{r_2-r_1}}r
\end{equation}
with $F^x_{in2},\dots,F^y_{out1}$ constants. (As in the previous example, $[F]=[L]^{1/2}$.) For the ingoing cases (Figure 1), $r_2=0\vert_+$, $F^x_{in2}=0$ (since $\Theta^x_{in}(0\vert_+)=-\infty)$) and $r_1=2M$, while for the outgoing cases (Figure 2), $r_1=0\vert_+$, $F^y_{in1}=0$ (since $\Theta^y_{out}(0\vert_-)=+\infty)$) and $r_2=2M$ .

\

The Lagrangians ${\cal L}={{1}\over{2}}\dot{F}^2$ and actions $S=\int d\lambda{\cal L}$ lead to the Feynman propagators $K^x_{in}(F^x_{in2},r_2;F^x_{in1},r_1)$ and  

$K^y_{out}(F^y_{out2},r_2;F^y_{out1},r_1)$ given by 
\begin{equation}
K^x_{in}(H^\pm\buildrel{B.H.}\over\longrightarrow 0\vert_+)=K^x_{in}(0,0\vert_+;F^x_{in1},2M)={{e^{{{i}\over{4}}(\pi-(F^x_{in1})^2/M)}}\over{\sqrt{4\pi M}}} 
\end{equation}
and 
\begin{equation}
K^y_{out}(0\vert_-\buildrel{W.H.}\over\longrightarrow H^\mp)=K^y_{out}(F^y_{out2},2M;0,0\vert_-)={{e^{-{{i}\over{4}}(\pi-(F^y_{out2})^2/M)}}\over{\sqrt{4\pi M}}} 
\end{equation}
which are finite.
\section{Final comments}
We have shown that to any affinely parametrized geodesic congruence described by the Raychaudhuri equation, it can be assigned a Feynman propagator describing its flow. In particular, at the end of Subsection 5.2 we showed that, for the example defined in Subsection 5.1, the propagator remains {\it finite} at a caustic, where the classical expansion diverges.  This result could be valid in general, and is an indication that the introduction of a quantum description should smooth or even disappear the divergences (singularities) of the classical theory. (An example is the disappearance of the Schwarzschild black hole singularity in the context of loop quantum gravity [17].)

\

In the non-geodesic case (timelike or null), $v\cdot D(v^\mu)\neq 0$ and equation (3) becomes 
\begin{equation}
{{d\Theta}\over{d\lambda}}=-{{1}\over{n}}\Theta^2-\sigma_{\mu\nu}\sigma^{\mu\nu}+\omega_{\mu\nu}\omega^{\mu\nu}-R_{\mu\nu}v^\mu v^\nu+D\cdot a
\end{equation}
where $D\cdot a=a^\mu_{;\mu}$ and 
\begin{equation}
a^\mu(\lambda)=v^\nu D_\nu v^\mu={{dx^\nu}\over{d\lambda}}D_\nu v^\mu\equiv {{Dv^\mu}\over{d\lambda}} 
\end{equation}
is an acceleration term [2,18]. The introduction of the function $F(\lambda)$ as in (7) leads to an equation analogous to (8), namely
\begin{equation}
\ddot{F}(\lambda)+{\Omega^\prime}^2(\lambda)F(\lambda)=0
\end{equation}
with a modified frequency
\begin{equation}
{\Omega^\prime}^2={{1}\over{n}}(\sigma^2-\omega^2+R_{\mu\nu}v^\mu v^\nu)-a^\mu_{;\mu},
\end{equation} 
which again defines an oscillator with $\lambda$-time dependent frequency. Then to an arbitrary congruence of curves $x^\mu(\lambda)$, geodesic or non-geodesic, with enough differentiability in a spacetime defined by $g_{\mu\nu}$, $R_{\mu\nu}$, etc., can be associated, in principle, a quantum propagator along its flow. 

\

As a final remark, we wish to mention that closely related to the Raychaudhuri equation is the {\it geodesic deviation equation} (G.D.E.) which is the Jacobi equation for the deviation vector $\xi^\mu={{\partial x^\mu}\over{\partial s}}$ (Jacobi field) measuring the separation between geodesics in the congruence, each geodesic labelled by the real parameter $s$ (in this case $x^\mu=x^\mu(\lambda,s)$):
\begin{equation}
{{D^2}\over{d\lambda^2}}\xi^\mu+R^\mu_{\lambda\sigma\rho}v^\lambda v^\rho\xi^\sigma=0
\end{equation}
(${{D}\over{d\lambda}}\equiv v\cdot D$), where ${{D^2}\over{d\lambda^2}}\xi^\mu\equiv A^\mu$ is the relative acceleration between geodesics ({\it tidal acceleration}). It can be easily verified that coordinates can be chosen such that $v\cdot \xi=0$, and that it holds $v\cdot D\xi^\mu=B^\mu_\nu\xi^\nu$, from which (63) is derived. Thus, it is the {\it covariant gradient of the geodesic velocity} $v^\mu_{;\nu}=B^\mu_\nu$ (eq.(4)) the basic quantity from which both the Raychaudhuri equation and the G.D.E. emerge. Finally, it is clear that at a caustic $\xi^\alpha$ vanishes, and that in flat spacetime $A^\mu=0$ i.e. tidal forces are pure geometrical phenomena.
\section*{Acknowledgments}
The author thanks Oscar Brauer for the realization of the Figures, and to an anonymous Referee for his suggestions and remarks to improve the manuscript.
\section*{References}

\

[1] Raychaudhuri, A. Relativistic Cosmology. I, Physical Review $\bf{98}$, 1123-1126, (1955).

\

[2] Kar, S. and Sengupta, S. The Raychaudhuri equations: A brief review, Pramana $\bf{69}$, 49-76, (2007).

\

[3] Tipler, F.J. General Relativity and Conjugate Ordinary Differential Equations, Journal of Differential Equations $\bf{30}$, 165-174, (1978).

\

[4] Teschl, G. {\it Ordinary Differential Equations and Dynamical Systems}, Providence: American Mathematical Society, (2012).

\

[5] Feynman, R.P. and Hibbs, A.R. {\it Quantum Mechanics and Path Integrals}, McGraw-Hill, N.Y., (1965). 

\

[6] Schulman, L.S. {\it Techniques and Applications of Path Integration}, Wiley, N.Y., (1981).

\

[7] Kruskal, M.D. Maximal extension of Schwarzschild metric, Physical Review $\bf{119}$, 1743-1745, (1960).

\

[8] Szekeres, G. On the singularities of a Riemannian manifold, Publ. Math. Debrecen $\bf{7}$, 285-301, (1960).

\

[9] Ehlers, J. The Nature and Structure of Spacetime, in {\it The Physicist's Conception of Nature}, J. Mehra (ed.), D. Reidel Pub. Co., Dordrecht-Holland, 71-91, (1973).

\

[10] Rovelli, C. {\it Quantum Gravity}, Cambridge University Press, pbk. ed. (2008).

\

[11] Álvarez, E. Windows on Quantum Gravity, Fortschritte der Physik $\bf{69}$, 2000080, 1-29, (2021).

\

[12] Baez, J.C. and Bunn, E.F. The meaning of Einstein equation, American Journal of Physics $\bf{73}$, 644-652, (2005).

\

[13] Gradshtein, I.S. and Ryzhik, I.M. {\it Table of Integrals, Series, and Products}, Academic Press, (1980).

\

[14] Arfken, G.B. and Weber, H.J. {\it Mathematical Methods for Physicists}, Academic Press, San Diego, (2001).

\

[15] Poisson, E. {\it A Relativistic Toolkit, The Mathematics of Black-Hole Mechanics}, Cambridge University Press, (2004).

\

[16] Socolovsky, M. Eikonal Equations for Null Radial Geodesics in the Schwarzschild Metric, Theoretical Physics $\bf{5}$, 41-49, (2020).

\

[17] Modesto, L. Disappearance of the black hole singularity in loop quantum gravity, Physical Review D $\bf{70}$, 124009, 1-5, (2004).

\

[18] Senovilla, J.M.M. and Garfinkle, D. The 1965 Penrose singularity theorem, Classical and Quantum Gravity $\bf{32}$, 124008, 1-45, (2015).
\end{document}